\documentclass[11pt,DIV=12]{scrartcl}
\pdfoutput=1

\makeatletter
\DeclareOldFontCommand{\rm}{\normalfont\rmfamily}{\mathrm}
\DeclareOldFontCommand{\sf}{\normalfont\sffamily}{\mathsf}
\DeclareOldFontCommand{\tt}{\normalfont\ttfamily}{\mathtt}
\DeclareOldFontCommand{\bf}{\normalfont\bfseries}{\mathbf}
\DeclareOldFontCommand{\it}{\normalfont\itshape}{\mathit}
\DeclareOldFontCommand{\sl}{\normalfont\slshape}{\@nomath\sl}

\usepackage[T1]{fontenc}
\usepackage[utf8]{inputenc}
\usepackage{lmodern}
\usepackage{amsmath}
\usepackage{amsfonts}
\usepackage{amssymb}
\usepackage[protrusion=true,expansion,kerning=true,tracking=true,final]{microtype}
\usepackage{bbold}
\usepackage{graphicx}
\usepackage[titletoc,title]{appendix}

\usepackage[usenames,dvipsnames,table]{xcolor}
     \definecolor{hgreen}{rgb}{0,.3,0}
     \definecolor{hred}{rgb}{.3,0,0}
     \definecolor{hblue}{rgb}{0,0,.3}
     \definecolor{LightGray}{gray}{0.95}
     \definecolor{DarkerGray}{gray}{0.80}
     \definecolor{MathBlue}{rgb}{0.368417, 0.506779, 0.709798}
     \definecolor{MathYellow}{rgb}{0.880722, 0.611041, 0.142051}
     \definecolor{MathGreen}{rgb}{0.560181, 0.691569, 0.194885}
     \definecolor{MathRed}{rgb}{0.922526, 0.385626, 0.209179}
     \definecolor{MathViolet}{rgb}{0.528488, 0.470624, 0.701351}
\usepackage{syntonly}
\usepackage{slashed}
\usepackage[font=small,format=plain,labelfont=bf,up,textfont=it,up,width=.9\textwidth]{caption}
\usepackage{subfig}
\usepackage{array}
\usepackage[colorlinks=true,
	    linkcolor=hblue,
	    citecolor=hgreen,
	    filecolor=hblue,
	    urlcolor=RoyalBlue]{hyperref}
\usepackage{comment}
\usepackage[numbers,sort&compress]{natbib}
\usepackage[auth-lg,affil-sl]{authblk}
\usepackage{fancyhdr}

\usepackage{tikz}
\usetikzlibrary{shapes.geometric, arrows, positioning}
\tikzstyle{rect} = [rectangle, text centered, draw=black, fill=MathBlue!60!]
\tikzstyle{input} = [rectangle, rounded corners, text centered, draw=black, fill=MathRed!60!]
\tikzstyle{output} = [ellipse, rounded corners, text centered, draw=black, fill=MathYellow!60!]
\tikzstyle{arrow} = [thick,->,>=stealth]

\usepackage{listings}
\definecolor{darkgray}{rgb}{0.95,0.95,0.95}
\definecolor{comment}{rgb}{0.4,0.4,0.4}
\newcommand{\MaRTIn}{\texttt{MaRTIn}}
\newcommand{\foldface}[1]{{\footnotesize\textsf{#1}}}

\fancypagestyle{plain}{
  \fancyhead{}
  \fancyfoot{}

}

\KOMAoptions{
parskip=half,
abstract=true,
numbers=noenddot,
DIV=12
}

\begin{document}
\renewcommand\Authands{, }
\setcounter{page}{1}
\pagenumbering{roman}

\title{
\rule{\linewidth}{2pt}
\Huge\MaRTIn{}}
\subtitle{\Large Manual for the ``Massive Recursive Tensor Integration''\\
\rule{\linewidth}{2pt}\\[2em]}
\date{\large\today}
\author[a]{Joachim Brod%
        \thanks{\texttt{joachim.brod@uc.edu}}}

\author[b]{Lorenz Hüdepohl%
        \thanks{\texttt{lorenz.huedepohl@mpcdf.mpg.de}}}

\author[c]{Emmanuel Stamou%
        \thanks{\texttt{emmanuel.stamou@tu-dortmund.de}}}

\author[a,c]{Tom Steudtner%
        \thanks{\texttt{tom2.steudtner@tu-dortmund.de}}}

\affil[a]{{Department of Physics, University of Cincinnati, Cincinnati, OH 45221, USA}}
\affil[b]{{Max Planck Computing and Data Facility, D-85748 Garching, Germany}}
\affil[c]{{Department of Physics, TU Dortmund, D-44221 Dortmund, Germany}}

\maketitle

\begin{abstract}
  \normalsize We present \MaRTIn{}, an extendable all-in-one package
  for calculating amplitudes up to two loops in an expansion in
  external momenta or using the method of infrared
  rearrangement. Renormalizable and non-renormalizable models can be
  supplied by the user; an implementation of the Standard Model is
  included in the package. In this manual, we discuss the scope and
  functionality of the software, and give instructions of its use.
\end{abstract}

\clearpage
\pdfbookmark[1]{Table of Contents}{tableofcontents}
\setcounter{page}{1}
\tableofcontents
\thispagestyle{empty}
\clearpage

\setcounter{page}{1}
\pagenumbering{arabic}

\section{Overview\label{sec:Overview}}

Perturbation theory is one of the most reliable tools to obtain quality
predictions from quantum field theories. In order to further enhance the
precision, higher loop orders are required, increasing the computational
complexity of the calculation and with it the need for more sophisticated
software pipelines to automatise them. Traditionally, these tool chains follow
a modular structure, though not all pieces or build scripts are public. For
instance, generic Feynman diagrams can be generated via
\texttt{QGRAF}~\cite{Nogueira:1991ex},
\texttt{Feynarts}~\cite{Hahn:1998yk,Hahn:2000kx} or future versions of
\texttt{FORM}~\cite{Vermaseren:2000nd, Kuipers:2012rf}. These expressions
require further manipulation such as inserting Feynman rules, problem-specific
diagram filtering and expansions, partial fractioning, tensor reduction and
topology identification. Subsets of these tasks can be delegated to tools like
\texttt{DIANA}~\cite{Tentyukov:1999is}, \texttt{q2e} and
\texttt{exp}~\cite{Harlander:1998cmq,Seidensticker:1999bb},
\texttt{LIMIT}~\cite{Herren:2020ccq}, \texttt{tapir}~\cite{Gerlach:2022qnc},
\texttt{TopoID}~\cite{Hoff:2016pot}, \texttt{Alibrary}~\cite{Alibrary} and
\texttt{Feynson}~\cite{Feynson}. Moreover, the algebra of spinor contractions
and symmetry groups has to be simplified. Note that there is a plethora of
software packages automatising computations at tree-level and one-loop order,
employing a Passarino--Veltman decomposition of
integrals~\cite{Passarino:1978jh}. However, already at two loops a more
sophisticated methodology is required. Integration-by-parts (IBP) relations are
recursively applied to loop integrals until they have reduced to master
integrals. This step may be performed by
\texttt{LiteRed}~\cite{Lee:2012cn,Lee:2013mka},
\texttt{FIRE}~\cite{Smirnov:2008iw,Smirnov:2013dia,Smirnov:2014hma,Smirnov:2019qkx},
\texttt{Reduze}~\cite{Studerus:2009ye,vonManteuffel:2012np} or
\texttt{Kira}~\cite{Maierhofer:2017gsa,Klappert:2020nbg}. Examples of packages
combining the IBP reductions and insertion of master integrals for specific
families of problems are \texttt{MINCER}~\cite{Larin:1991fz},
\texttt{MATAD}~\cite{Steinhauser:2000ry}, \texttt{FORCER}~\cite{Ruijl:2017cxj}
and \texttt{FMFT}~\cite{Pikelner:2017tgv}.

\MaRTIn{} (Massive Recursive Tensor Integration) aims to be an
all-in-one tool for the computation of multi-leg amplitudes in an
expansion of small external momenta at tree and loop level in
arbitrary quantum field theories. A typical computation starts from
Feynman rules specified by the user, which may comprise both
renormalisable and non-renormalisable operators. \MaRTIn{} leverages
\texttt{QGRAF}~\cite{Nogueira:1991ex} to generate Feynman diagrams
automatically, and proceeds with the main computation using routines
implemented in \texttt{FORM}~\cite{Vermaseren:2000nd,Kuipers:2012rf}. This includes
the resolution of spinor, gauge, and global symmetry index
contractions, as well as the tensor and IBP reduction and, eventually,
scalar loop integration. Several schemes for the treatment $\gamma_5$
are available.

The major mode of operation employs either an expansion in all
external momenta, or the infrared rearrangement outlined in
Refs.~\cite{Misiak:1994zw, Chetyrkin:1997fm}. In either case,
integrals are reduced to vacuum topologies with arbitrary, possibly
vanishing mass arguments. The integration is then performed
symbolically up to two-loop order in $4-2\varepsilon$ space-time
dimensions~\cite{Davydychev:1992mt,Bobeth:1999mk}, to a user-specified
order in powers of $\varepsilon$ and external momenta. The
functionality to perform calculations in $3-2\varepsilon$ space-time
dimensions is contained in the package, but is not as well-tested.

Another mode of operation retains the full momentum dependence but does not
perform any integration. This is implemented up to one-loop order. The
functionality of \MaRTIn{} may be extended in future releases.

\MaRTIn{} produces outputs in both \texttt{FORM} and
\texttt{Mathematica}~\cite{Mathematica} compatible formats, and optionally a
graphical representation of all involved Feynman diagrams.

In the following sections, we specify in detail how \MaRTIn{} is
obtained, installed, operated, and extended.

\section{Setup\label{sec:Setup}}

\MaRTIn{} is distributed under the \textit{GNU Public License, Version
  3}~\cite{gplv3} and available to download at
\href{https://gitlab.com/manstam/martin}{{\color{RoyalBlue}
    \texttt{https://gitlab.com/manstam/martin}}}. The code itself is
mostly written in and tested against
\texttt{FORM}~4~\cite{Kuipers:2012rf}, which has to be installed
separately. Similarly, \MaRTIn{} requires
\texttt{QGRAF}~\cite{Nogueira:1991ex} to generate diagrams. Note that
both packages have several subdependencies and need to be compiled
from their source codes. Thus, a POSIX compatible system with a
working GNU toolchain is required. \MaRTIn{} is operated using
\texttt{GNU make} as well as a \texttt{python3} script, which in turn
requires the common modules \texttt{os}, \texttt{sys},
\texttt{shutil}, \texttt{re}, \texttt{argparse}, \texttt{subprocess}
and \texttt{configparser}. The underlying Makefile also utilises
\texttt{bash}, (GNU) \texttt{awk} and several other standard shell
programs.  In order to output Feynman diagrams in PDF format,
\texttt{graphviz} with the \texttt{neato} engine are
required. Optionally, the package \texttt{richard\_draw} may be
utilised, which produces higher-quality output using
\texttt{lualatex}. It can be downloaded at
\href{https://gitlab.com/manstam/richard_draw}{{\color{RoyalBlue}
    \texttt{https://gitlab.com/manstam/richard\_draw}}}.

Once all prerequisites are met, \MaRTIn{} needs to be set up in three
places:
\begin{itemize}
\item The source code directory \texttt{martin}.  It requires no
  further installation, but a system soft link to the main executable
  \texttt{martin/martin.py} is recommended. The source directory
  includes the subdirectories \texttt{code} with the source codes,
  \texttt{user\_template} with some prototypes of the files, and
  further directories mentioned below, as well as this manual.
\item A config file \texttt{.martin.conf}, which should be manually
  placed in the user's home directory. A template version of the file
  is contained in \newline
  {\texttt{martin/user\_template/template\_martin.conf}}. This file
  specifies the locations to the \texttt{FORM} and \texttt{QGRAF}
  executables, the number of default parallel processes, \texttt{FORM} options
  and default paths. The options are detailed in the template file
  itself.
\item The user's working directory, e.g., named \texttt{martin\_user},
  which should be created by the
  user and populated with the following subdirectories \\
  (examples are found in \texttt{martin/user\_template}):
  \begin{itemize}
  \item \texttt{models}: This directory contains all user-defined
    model files. Further details are deferred to Sec.~\ref{sec:Model}.
  \item \texttt{problems}: contains (possibly nested) subdirectories,
    which can have arbitrary names. Each of these subdirectories may
    contain several \texttt{loop.*.dat} files. Every such file
    specifies all options required for a single run of \MaRTIn. The
    contents of these files will be explained in Sec.~\ref{sec:Workflow}.
  \item \texttt{prc}: is the canonical location for user-defined
    \texttt{FORM} routines that will be automatically detected by
    \MaRTIn. The routines can then be included in the \texttt{FORM}
    folds of the model or \texttt{loop.*.dat} files
    (see Sec.~\ref{sec:problem:file}).
  \item \texttt{results}: This is the location were \MaRTIn{}
    automatically writes intermediate and final output, creating the
    same folder structure as in the user-defined \texttt{problems}
    directory. The user should not manually modify or add to the
    contents of this directory since \MaRTIn{}'s functionality can
    overwrite and delete these directories.
  \end{itemize}
\end{itemize}

In order to test whether the \MaRTIn{} installation is complete, users
can calculate the example described in greater detail in
Sec.~\ref{sec:Examples}. All files necessary for this run are
contained in \MaRTIn{}.  Assuming that \texttt{FORM}, \texttt{QGRAF},
and \MaRTIn{} are properly set up, the command
\begin{lstlisting}
martin ./user_template/problems/SM/loop.1_uu.dat
\end{lstlisting}
invoked from the main \MaRTIn{} directory should generate and compute
the diagram in Fig.~\ref{fig:rdraw_example}. The command
\begin{lstlisting}
martin ./user_template/problems/SM/loop.1_uu.dat pdf
\end{lstlisting}
should generate a pdf file of the corresponding diagrams. Using
\texttt{rpdf} instead of the target \texttt{pdf} would instead use
\texttt{richard\_draw} to create a pdf file for the diagrams. A
successful run will provide the screen printout discussed in
Sec.~\ref{sec:Examples}, and save all results in the directory
\texttt{./user\_template/results/SM/}.

In the next section, we describe the concrete workflow of \MaRTIn{} in
detail.

\section{Workflow\label{sec:Workflow}}

\begin{table}
  \centering
  \begin{tabular}{|l|l|}
    \rowcolor{DarkerGray} \hline Input files & Output files \\
    \hline
    \hspace{-0.8em}\begin{tabular}{l}
    \texttt{QGRAF} files (diagrams) \\
    \quad \texttt{models/qgraf/\textit{model}.prop.lag}  \\
    \quad \texttt{models/qgraf/\textit{model}.vrtx.lag}  \\[.5em]
    \texttt{FORM} model file (Feynman rules) \\
    \quad \texttt{models/form/model\_\textit{MODEL}} \\[.5em]
    Problem file (process and options)\\
    \quad \texttt{problems/\textit{process}/loop.\textit{problem}.dat} \\[.5em]
    Optional: \texttt{richard\_draw} file (drawing)\\
    \quad \texttt{models/rdraw/\textit{model}.json}
    \end{tabular}\hspace{-.8em} &
    \hspace{-.8em}\begin{tabular}{l}
      \texttt{FORM} output \\
      \quad \texttt{results/\textit{process}/form.\textit{problem}/dia*.sav}\\[.5em]
      \texttt{Mathematica} output \\
      \quad \texttt{results/\textit{process}/math.\textit{problem}/dia*.m} \\[.5em]
      Feynman graphs \\
      \quad \texttt{results/\textit{process}/graphs.\textit{problem}.pdf} \\[.5em]
      \texttt{richard\_draw} Feynman graphs \\
      \quad \texttt{results/\textit{process}/rgraphs.\textit{problem}.pdf} \\[.5em]
    \end{tabular}\hspace{-.8em} \\
    \hline
  \end{tabular}
  \caption{Overview of \MaRTIn{} input and output files. See text for
    details.}
  \label{tab:inoutput}
\end{table}

In this section we describe how \MaRTIn{} is used.
A list of the
required input files, as well as the output files, is contained in
Tab.~\ref{tab:inoutput}. The two key elements for the operation of
\MaRTIn{} are the \texttt{loop.\textit{problem}.dat}, which contains
all information about the calculation at hand, as well as the main
executable \texttt{martin.py}, which serves as a convenience wrapper
around the Makefile. Note that \texttt{\textit{problem}} here
represents a placeholder for a user-defined identifier.  Different
steps of the computation are triggered by \texttt{\textit{targets}}
arguments passed to the executable. Assuming that the full path to
\texttt{martin.py} has been aliased as the command \texttt{martin}, a
typical \MaRTIn{} call has the following shape:
\begin{lstlisting}
  martin problems/@\textit{process}@/loop.@\textit{problem}@.dat @\textit{targets}@
\end{lstlisting}
The command
\begin{lstlisting}
  martin -h
\end{lstlisting}
lists and explains all target options.
There are many targets that
are only important for the inner workings of \MaRTIn{} and are
evaluated successively. By default, output of all targets (including
intermediate ones) are written to the directory
\texttt{results/\textit{process}/}. For the user, the most relevant
targets are the following:
\begin{itemize}
\item \texttt{all} is the default target of \texttt{martin.py}
  and will be called if no target argument is specified. It computes
  all diagrams and saves the results in the \texttt{results} folder,
  as discussed above. For every diagram that has not been computed
  already, \MaRTIn{} will spawn a separate process to do so. The
  maximal number of parallel processes can be configured in
  \texttt{.martin.config}.
\item \texttt{diaNUM} computes the single diagram with number
  \texttt{NUM}. The result is written to \linebreak
  {{\texttt{results/\textit{process}/form.\textit{problem}/diaNUM.sav}}}
  using \texttt{FORM}s binary format, as well as \linebreak
  {{\texttt{results/\textit{process}/math.\textit{problem}/diaNUM.m}}}
  in Wolfram Language.
\item \texttt{sum} computes the sum of all Feynman diagrams. The
  results are stored as \linebreak {\centering
    {\texttt{results/\textit{process}/form.\textit{problem}/diasum.sav}}}
  in \texttt{FORM} as well as \newline
  {{\texttt{results/\textit{process}/math.\textit{problem}/diasum.m}}}
  in \texttt{Mathematica} compatible format. If any individual
  \texttt{diaNUM.sav} files are missing, they will be computed with
  the \texttt{all} target.
\item \texttt{pdf} will generate a PDF file containing the graphical
  representation of each Feynman diagram. This is stored in
  {{\texttt{results/\textit{process}/graphs.\textit{problem}.pdf}}}.
\item \texttt{rpdf} utilises \texttt{richard\_draw} to generate a
  higher-quality pdf file containing all Feynman graphs. This is stored
  under {{\texttt{results/\textit{process}/rgraphs.\textit{problem}.pdf}}}.
\item \texttt{clean} removes all final and intermediate results of the
  selected problem.
\end{itemize}

The executable \texttt{martin.py} also takes arguments that can be used to override
options in the config file \texttt{.martin.conf}, call \texttt{martin -h} for further details.
In the remainder of this section, we
specify the content and structure of a valid \texttt{loop.\textit{problem}.dat} file,
before giving an overview of \MaRTIn{}'s algorithm.

\subsection{Problem file}\label{sec:problem:file}

The file \texttt{problems/\textit{process}/loop.\textit{problem}.dat}
contains all necessary information for a single run of \MaRTIn{}. It
specifies how the model is assembled from the modular pieces that will
be introduced in Sec.~\ref{sec:Model}, defines in detail which Feynman diagrams
are to be calculated, configures various aspects of the computation
itself, and specifies if and how user-defined code is injected.
Example files can be found in the \MaRTIn{} source directory at
\texttt{user\_template/problems/SM/}.

The file itself consists of
\texttt{FORM} folds, i.e. named objects of the syntax
\begin{lstlisting}
  *--#[ FOLDNAME :
    @\textit{some code here}@
  *--#] FOLDNAME :
\end{lstlisting}
which are being inserted at the right place by \texttt{FORM}'s
preprocessor. These objects must never be deleted from the file, even
if they are empty. The content of each fold is  \texttt{FORM}
code. The only exception is the very first fold we encounter, which
actually contains \texttt{QGRAF}'s own scripting language. In fact,
this fold contains the contents of the \texttt{QGRAF} input file:
\begin{lstlisting}
  *--#[ QGRAF :
  @\textcolor{comment}{* ~ This is a comment.}@
  @\textcolor{comment}{* ~ Do not remove or reorder statements below.}@

  @\textcolor{comment}{* ~ Define which QGRAF model files to use.}@
  @\textcolor{comment}{* ~ We typically split them into propagators and vertices, but that is not required.}@
  @\textcolor{comment}{* ~ Both are should be valid files located in models/qgraf/ }@
  @\textcolor{comment}{* ~ The functionality of including multiple models in this}@
  @\textcolor{comment}{* ~ way is a MaRTIn feature; it is not possible in solo QGRAF runs}@
    model = "@\textit{qmodel}@.prop.lag";
    model = "@\textit{qmodel}@.vrtx.lag";

  @\textcolor{comment}{* ~ Define external legs using field names. }@
  @\textcolor{comment}{* ~ Use q's for external momenta (note that martin substitutes q0 = 0. }@
    in  = @\textit{field1}@[q1], @\textit{field2}@[q2];
    out = @\textit{field3}@[q3], @\textit{field4}@[q4];

  @\textcolor{comment}{* ~ Loop order, e.g. 2. Loop momenta have to be p1, p2.}@
    loops = 2;
    loop_momentum = p;

  @\textcolor{comment}{* ~ Some options to select topologies, e.g.:}@
    options = onepi;

  @\textcolor{comment}{* ~ More selectors like vsums and psums here}@

  *--#] QGRAF :
\end{lstlisting}
Note that \MaRTIn{} allows for the presence of several separate
\texttt{QGRAF} model files; they are combined automatically before
being passed to \texttt{QGRAF} by concatenating (\texttt{cat}) them
in the order they appear in the fold.
It is the user's responsibility to
ensure that the model files are consistent with each other. We refer
to the \texttt{QGRAF} manual for a more detailed description of the
valid syntax.

The next fold sets up important options for the \texttt{FORM} part of
\MaRTIn{}. These are saved as preprocessor variables, which are
collected and described in Tab.~\ref{tab:fold-MAIN}. An example and more explanations are
given below.
\begin{table}
  \centering
  \begin{tabular}{|l|l|}
    \rowcolor{DarkerGray} \hline Option & Description \\
    \hline \texttt{NM} & Represents the number of \texttt{FORM} model files that are\\
    &  being combined  for the current problem.\\[0.5em]
    \rowcolor{darkgray} \texttt{MODEL\textit{i}} & Series of variables with integer \texttt{\textit{i}}
                                                   going from 1 to \texttt{NM}. \\
    \rowcolor{darkgray} &   Each refers to a filename \texttt{model\_MODELx} of a \\
    \rowcolor{darkgray} & \texttt{FORM} model file found in \texttt{models/form/}.\\[0.5em]
    \texttt{EXPDENO} & Before integration, expand in external momenta to a \\
    &   power given by this variable. This option is  \\
    &  exclusive with \texttt{IRA} and \texttt{GENERICLOOPFUNCTIONS}. \\[0.5em]
    \rowcolor{darkgray} \texttt{IRA} & Perform infrared rearrangement before integration.  \\
    \rowcolor{darkgray}  & Expand external momenta up to a power specified by \\
    \rowcolor{darkgray}  &this variable. Exclusive option with  \texttt{EXPDENO}  and \\
    \rowcolor{darkgray}  & \texttt{GENERICLOOPFUNCTIONS}. \\[0.5em]
    \texttt{GENERICLOOPFUNCTIONS} & Skip integration, collect integrand in generic function. \\
    & Exclusive option with \texttt{EXPDENO}  and \texttt{IRA}. \\[0.5em]
    \rowcolor{darkgray}\texttt{FINALEPLIM} & Power in $\varepsilon$ up to which the final result is expanded. \\[0.5em]
    \texttt{DSCHEME} & Scheme for handling $\gamma_5$. Options are \texttt{"NDR"} (default), \\
    & \texttt{"sNDR"}, \texttt{"HV"} and \texttt{"LARIN"}. See main text for details.\\[0.5em]
    \rowcolor{darkgray}\texttt{CLLABEL} & If defined, mark each closed fermion loop with a function \\
    \rowcolor{darkgray} & \texttt{cl} tracking the fields in the loop. \\[0.5em]
    \texttt{PRINT} & If defined, intermediate steps in the computation of \\
    & each diagram are printed to the terminal. \\ [0.5em]
    \rowcolor{darkgray} \texttt{FINALPRINT} & If defined, the final result of each diagram is printed \\
    \rowcolor{darkgray} & to the terminal. \\ \hline
  \end{tabular}
  \caption{
  Overview of preprocessor options to configure \MaRTIn{}. These are found in the fold \foldface{MAIN} located in each problem file. }
  \label{tab:fold-MAIN}
\end{table}
\begin{lstlisting}
  *--#[ MAIN :
  @\textcolor{comment}{* ~ This is a FORM comment.}@

  @\textcolor{comment}{* ~ Each model may consist of several form model files found in models/form/ }@
  @\textcolor{comment}{* ~ The total number of model files must be given as NM. }@
  @\textcolor{comment}{* ~ For instance, for two model files with names ``model\_fmodelA'' and ``model\_fmodelB'': }@
    #define NM "2"
    #define MODEL1 "@\textit{fmodelA}@"
    #define MODEL2 "@\textit{fmodelB}@"

  @\textcolor{comment}{*** Modes of operation. Only one of these three should be defined. }@
  @\textcolor{comment}{* ~ Expand in external momenta to this power }@
    #define EXPDENO "2"
  @\textcolor{comment}{* ~ Define if infrared rearrangement should be used.}@
    #define IRA "1"
  @\textcolor{comment}{* ~ No integration, keep a generic expression of unsolved integrals}@
    #define GENERICLOOPFUNCTIONS
\end{lstlisting}

Defining \texttt{EXPDENO "$n$"} will perform a Taylor expansion of all
propagators $\frac{1}{(p+q)^2 - m^2}$ up to powers $n$ in all external
momenta $q$. Alternatively, setting the option \texttt{IRA ``$n$''}
the code performs an infrared rearrangement~\cite{Misiak:1994zw,
Chetyrkin:1997fm}, i.e., it introduces the regularising mass
parameter $m_\text{IRA}$ via the repeated application of the identity
\begin{equation}
  \frac{1}{(p + q)^2 - M^2}
 = \frac{1}{p^2 - m_\text{IRA}^2}
  + \frac{ M^2 - m_\text{IRA}^2 - 2 \,p\!\cdot\! q - q^2}{p^2 - m_\text{IRA}^2}
     \frac{1}{(p + q)^2 - M^2} \,.
\end{equation}
External momenta are kept up to power $n$. In practice, this identity
is applied recursively to every term until its degree of divergence is
negative; these terms are then dropped. Note that the $m_\text{IRA}^2$ in
the numerator of the second term on the right side of the equation is
dropped; accordingly, additional local counterterms must be introduced
(see Ref.~\cite{Chetyrkin:1997fm} for details).

\begin{lstlisting}
  @\textcolor{comment}{* ~ Order in $\varepsilon$ to which final result is expanded}@
  @\textcolor{comment}{* ~ This is not relevant for the GENERICLOOPFUNCTIONS option}@
    #define FINALEPLIM "-1"

  @\textcolor{comment}{*** Scheme how gamma5 should be treated. The options are: }@
  @\textcolor{comment}{* ~ NDR       -- naive dimensional regularisation, abort on gamma5 encounter inside traces}@
  @\textcolor{comment}{* ~ sNDR      --  semi-naive dimensional regularisation, use naive relations }@
  @\textcolor{comment}{* ~ \phantom{sNDR   -- } in $(4-2\varepsilon)$ dimensions, but set flagsNDR }@
  @\textcolor{comment}{* ~ HV        -- Breitenlohner - Maison - 't Hooft - Veltman treatment }@
  @\textcolor{comment}{* ~ LARIN     -- Larin's scheme: replace gamma5 by Levi-Civita contraction }@
    #define DSCHEME "NDR"

\end{lstlisting}

A few comments about the implemented $\gamma_5$ schemes are in
order. The issues with $\gamma_5$ arise due to the choice of
dimensional regularisation~\cite{Jegerlehner:2000dz}. The $\gamma_5$
matrix is defined as
\begin{equation}\label{eq:gamma_5-def}
  \gamma_5 = \frac{i}{4!} \epsilon_{\mu \nu \rho \sigma}\, \gamma^\mu \gamma^\nu \gamma^\rho \gamma^\sigma \,,
\end{equation}
where $\epsilon_{\mu \nu \rho \sigma}$ denotes the completely
antisymmetric Levi-Civita tensor in four space-time dimensions, with
the convention $\epsilon_{0123} = 1$.

It is well known that the anticommutation relation
\begin{equation}\label{eq:gamma_5-anticommutation}
\{\gamma_5, \gamma^\mu\} = 0
\end{equation}
is algebraically inconsistent with the trace operation in $d \neq 4$
space-time dimensions~\cite{Collins:1984xc}.  In order enable
consistent treatments of $\gamma_5$, \MaRTIn{} internally can keep
track of whether fermion lines are open or closed, how many $\gamma_5$
appear, and whether Lorentz indices are $(4-2\varepsilon)$-, $4$- or
$(-2 \varepsilon)$-dimensional. The following options to handle
$\gamma_5$ are currently implemented:
\begin{itemize}
\item \texttt{\#define DSCHEME "NDR"} : ``Naive dimensional regularisation''\\
  This is the default choice of the algorithm.
  If $\gamma_5$ is encountered inside a trace, \MaRTIn{} exits with an error since NDR
  can give algebraically inconsistent results in traces.
  In open fermion lines, $\gamma_5$ matrices are moved to the right,
  assuming anticommutation $\{\gamma_5, \gamma^\mu\} = 0$.
\item \texttt{\#define DSCHEME "sNDR"} : ``Semi-Naive dimensional regularisation''~\cite{Chetyrkin:2012rz,Bednyakov:2012en}\\
  This algorithm always uses the anticommutation relation in
  Eq.~\eqref{eq:gamma_5-anticommutation}.  Inside traces, $\gamma_5$
  is replaced via the relation in Eq.~\eqref{eq:gamma_5-def}, and the
  Levi-Civita tensors and their contractions are formally treated as
  $(4-2\epsilon)$-dimensional. The implication of this simplifying
  assumption is that only $(4-2\epsilon)$-dimensional metric tensors
  and $\gamma$-matrices will appear in the final result.  Of course,
  the relation in Eq.~\eqref{eq:gamma_5-def} is then actually
  inconsistent unless additional terms $\propto \varepsilon$ are
  introduced, which this scheme neglects. Instead, whenever this
  insertion is made the term is marked by the variable
  \texttt{flagsNDR}.  Thus, starting from a certain order of
  $\varepsilon$, the \texttt{sNDR} algorithm does not yield a
  consistent result. The highest consistent order is the first one at
  which \texttt{flagsNDR} appears. Tracking the \texttt{flagsNDR}
  dependence is the responsibility of the user.
\item \texttt{\#define DSCHEME "HV"} : ``'t~Hooft--Veltman scheme''~\cite{tHooft:1972tcz,
    Breitenlohner:1975hg, Breitenlohner:1976te, Breitenlohner:1977hr}\\
  The $\gamma_5$ matrix is assumed to
  anticommute with $\gamma^\mu$ if $\mu = 0,1,2,3$, and to commute
  with $\gamma^\mu$, otherwise; i.e.
  $\left\{\gamma_5,\,\tilde \gamma^{\mu} \right\} = 0$ and
  $\left[\gamma_5,\,\hat \gamma^{\mu} \right] = 0$, where
  $\tilde \gamma^{\mu}$ and $\hat \gamma^{\mu}$ are the projections
  onto the $4$- and $d-4$-dimensional subspaces,
  respectively. Internally, projected Lorentz indices are contracted
  via the projected metric tensors \texttt{dd}, \texttt{ddtilde} and
  \texttt{ddhat} in $(4-2\varepsilon)$, $4$ and $(-2\varepsilon)$
  dimensions, respectively.
  Levi-Civita tensors are treated in four integer dimensions.
\item \texttt{\#define DSCHEME "LARIN"} : ``Larin scheme''~\cite{Larin:1993tq}\\
  In this scheme, any instance of $\gamma_5$ on both
  open and closed fermion lines  is replaced via the relation in
  Eq.~\eqref{eq:gamma_5-def} before traces are evaluated.
  The Levi-Civita tensor is treated in four integer dimensions.
\end{itemize}

\begin{lstlisting}

  @\textcolor{comment}{* ~ Define to introduce a label for each closed fermion line. }@
  @\textcolor{comment}{* ~ Syntax: cl(line\_id, field\_1 , ..., field\_n ) }@
    #define CLLABEL

  @\textcolor{comment}{* ~ Define to print the final result to the screen. }@
    #define FINALPRINT

  @\textcolor{comment}{* ~ Define to get more information about intermediate steps of the calculation. }@
    #define PRINT


  *--#] MAIN :
\end{lstlisting}
Finally, there are several folds to inject user-defined \texttt{FORM}
code, see also Sec.~\ref{sec:algo}.
\begin{lstlisting}
  *--#[ USERDEF :
  @\textcolor{comment}{* ~ Fill with your own FORM declarations.}@
  *--#] USERDEF :

  *--#[ TEST :
  @\textcolor{comment}{* ~ Code injection after model definitions, but before diagrams are generated.}@
  *--#] TEST :

  *--#[ FOLD1 :
  @\textcolor{comment}{* ~ Code injection right after expressions for each diagram are generated.}@
  *--#] FOLD1 :

  *--#[ FOLD2 :
  @\textcolor{comment}{* ~ Code injection before Dirac algebra treatment.}@
  *--#] FOLD2 :

  *--#[ FOLD3 :
  @\textcolor{comment}{* ~ Code injection before integration.}@
  *--#] FOLD3 :

  *--#[ FOLD4 :
  @\textcolor{comment}{* ~ Code injection after integration.}@
  *--#] FOLD4 :

  *--#[ FOLD5 :
  @\textcolor{comment}{* ~ Code injection after $\varepsilon$ expansion and coupling insertion.}@
  @\textcolor{comment}{* ~ Recommendation: replace function internally used for $(\mu/M)^{2\varepsilon}$. }@
    id eM(M?) = 1
              - 2*ep*Log(M)
              + 2*ep*Log(mu)
              + 2*ep^2*Log(M)^2
              + 2*ep^2*Log(mu)^2
              - 4*ep^2*Log(mu)*Log(M);
  *--#] FOLD5 :
\end{lstlisting}

In the next section, we detail the working algorithm of \MaRTIn{} and also
show where each fold is included.

\subsection{Algorithm\label{sec:algo}}

\begin{tikzpicture}[node distance=1.5cm and 1.5cm]
  \node (qgraf) [rect, align=center] {\texttt{QGRAF}\\ graph generation};
  \node (foldqgraf) [input, right=2cm of qgraf, align=center] {Fold \foldface{QGRAF} \\ in \texttt{loop.*.dat} file};
  \node (qgrafmodel) [input, right=1cm of foldqgraf, align=center] {\texttt{QGRAF} models};

  \node (dotfiles)  [output, below of=qgraf, align=center] {\texttt{.dot} files};
  \node (pdfgraphs)  [output, right=2cm of dotfiles, align=center] {\texttt{.pdf} graphs};

  \node (mainfrm) [rect, below of=dotfiles, align=center] {\texttt{main.frm:}\\ set up model};
  \node (mainfold)  [input, right=2.5cm of mainfrm, align=center] {Fold \foldface{MAIN}\\ in \texttt{loop.*.dat} file};
  \node (fmodel) [input, right=1.2cm of mainfold, align=center] {\texttt{FORM} models};
  \node (testfoldanchor) [circle, below=.5cm of mainfrm] {};
  \node (testfold)  [input, right=3.6cm of testfoldanchor, align=center] {Fold \foldface{TEST}\\ in \texttt{loop.*.dat} file};
  \node (gengraph) [rect, below=.5cm of testfoldanchor, align=center] {generate expressions\\insert Feynman rules \\
  (\texttt{gengraph})};
  \node (fold1anchor) [circle, below=.1cm of gengraph] {};
  \node (fold1)  [input, right=3.6cm of fold1anchor, align=center] {\foldface{FOLD1}\\ in \texttt{loop.*.dat} file};
  \node (expansion) [rect, below of=fold1anchor, align=center] { external momentum /  \\ IRA expansion };
  \node (dirac) [rect, below=.5cm of expansion, align=center] { $\gamma_5$ treatment  \\ spinor algebra };
  \node (fold2) [input, below=.3cm of fold1, align=center] { \foldface{FOLD2} \\ in \texttt{loop.*.dat} file};
  \node (info) [output, below=0.1cm of fold2, align=center] {  \texttt{.info} files \\ $\gamma_5$ details};
  \node (fold3anchor) [circle, below=.5cm of dirac] {};
  \node (fold3) [input, right=3.6cm of fold3anchor, align=center] { \foldface{FOLD3} \\ in \texttt{loop.*.dat} file};
  \node (integrate) [rect, below=0.5cm of fold3anchor, align=center] { tensor reduction  \\ IBP reduction \\ master integrals };
  \node (fold4anchor) [circle, below=0cm of integrate] {};
  \node (fold4) [input, right=3.6cm of fold4anchor, align=center] { \foldface{FOLD4} \\ in \texttt{loop.*.dat} file};
  \node (eps) [rect, below=0.2cm of fold4anchor, align=center] {  expand $\Gamma$s in $\varepsilon$ };
  \node (polanchor) [circle, below=0.2cm of eps] {};
  \node (pol) [input, right=3.6cm of polanchor, align=center] { \foldface{POLARIZATION}, \\ \foldface{INSERTCOUPLINGS} \\ in \texttt{FORM} models};
  \node (fold5anchor) [circle, below=0.cm of polanchor] {};
  \node (fold5) [input, below=1.8cm of fold4, align=center] { \foldface{FOLD5} \\ in \texttt{loop.*.dat} file};
  \node (finalexp) [rect, below=+0.2cm of fold5anchor, align=center] {final $\varepsilon$ expansion};
  \node (results) [output, below=0.5cm of finalexp, align=center] {final results\\ \texttt{.sav} files \\ \texttt{.m} files};

  \draw [arrow] (qgrafmodel) -- (foldqgraf);
  \draw [arrow] (foldqgraf) -- (qgraf);
  \draw [arrow] (qgraf) --  (dotfiles);
  \draw [arrow] (dotfiles) --  (mainfrm);
  \draw [arrow] (dotfiles) --  (pdfgraphs);
  \draw [arrow] (mainfrm) --  (gengraph);
  \draw [arrow] (fmodel) -- (mainfold);
  \draw [arrow] (mainfold) -- (mainfrm);
  \draw [arrow] (testfold) -- (testfoldanchor);
  \draw [arrow] (fold1) -- (fold1anchor);
  \draw [arrow] (gengraph) -- (expansion);
  \draw [arrow] (expansion) -- (dirac);
  \draw [arrow] (fold2) -- (dirac);
  \draw [arrow] (dirac) -- (info);
  \draw [arrow] (dirac) -- (integrate);
  \draw [arrow] (fold3) -- (fold3anchor);
  \draw [arrow] (fold4) -- (fold4anchor);
  \draw [arrow] (integrate) -- (eps);
  \draw [arrow] (eps) -- (finalexp);
  \draw [arrow] (finalexp) -- (results);
  \draw [arrow] (pol) -- (polanchor);
  \draw [arrow] (fold5) -- (fold5anchor);

\end{tikzpicture}

Simplified overview of \MaRTIn{}'s computational pipeline. Red rounded
squares represent input data, blue squares parts of the computation,
and yellow ellipses indicate output files.

\section{Model Definition\label{sec:Model}}

In the previous section, we saw that each model is defined by a
set of related \texttt{QGRAF} and \texttt{FORM} model files. In this
section, we will give some details on the vital components of these
files, which should enable users to implement their own models.

\MaRTIn{} uses the \texttt{QGRAF} and \texttt{FORM} model files to
assemble a symbolic representation of the amplitude from the
implemented Feynman rules. To achieve this, it uses the leg numbers
generated by \texttt{QGRAF} to (i) bring spinorial terms into the
correct order, (ii) generate all necessary (gauge and Lorentz)
indices, and (iii) substitute all Feynman rules.

\subsection{QGRAF files}

The \texttt{QGRAF}-model files are located in the working directory at
\texttt{models/qgraf/}. They are simply \texttt{QGRAF} model files,
the syntax may include any elements valid for the locally installed
\texttt{QGRAF} version. It is recommended to provide two separate
files, one for all propagators and one for all vertices, named
\texttt{\textit{model}.prop.lag} and \texttt{\textit{model}.vrtx.lag},
respectively.  The reason for this is that propagators must be defined
before vertices in \texttt{QGRAF}. Splitting each model in a
propagator and a vertex part makes it possible to include multiple
models. \MaRTIn{} will then automatically combine all \texttt{QGRAF}
model files into a single file in the correct order and pass it to
\texttt{QGRAF}. The propagator file essentially contains several
statements of the type
\begin{lstlisting}
  [@\textit{Boson}@, @\textit{BosonBar}@, +; pfunct='@\textit{function}@', m='@\textit{mass}@']

  [@\textit{Fermion}@, @\textit{FermionBar}@, -; pfunct='@\textit{function}@', m='@\textit{mass}@']
\end{lstlisting}
where cursive words are placeholders for identifiers to be defined by
the user.  To be precise, the first two argument define the names of
the fields and their adjoints. For real fields, both are the same.
The names must be declared symbols in the \texttt{FORM} model file.
The third argument determines if the field is fermionic or
bosonic. The keyword \texttt{pfunct} after the semicolon is required.
Its argument has to be declared as a commuting \texttt{FORM} function
in the \texttt{FORM} model file and represents the propagator
function. \MaRTIn{} automatically populates its arguments with
information about the propagator and its fields.  Similarly, the
keyword \texttt{m} is mandatory and represents the mass of the
propagating field.  It has to be a valid \texttt{FORM} symbol, to be
declared in the \texttt{FORM} model file. Note that \texttt{M0} is
implicitly assumed to be a vanishing mass. All additional keywords,
including \texttt{external}, are admissible.  In the same manner,
vertex file contains several statements of the form
\begin{lstlisting}
  [@\textit{Field1}@, @\textit{Field2}@, @\textit{Field3}@, @\textit{...}@; vfunct='@\textit{function}@']
\end{lstlisting}
where additional keywords are allowed. \texttt{vfunct} is
required by \MaRTIn{} and represents the vertex function. It has to be
declared in the \texttt{FORM} model file as a commuting function.
Note that \texttt{MaRTIn} provides its own
\texttt{QGRAF} style file. The output of \texttt{QGRAF} is distributed
into one \texttt{.dot} file per diagram, which are read by the
\texttt{FORM} routine and handled in a parallelised fashion.

\subsection{FORM files}

The \texttt{FORM} model files contain the main information
about the QFT at hand. While the \texttt{QGRAF} files define fields
and specify vertices by their legs, the actual Feynman rules are
implemented in the \texttt{FORM} model files, which \MaRTIn{}'s
\texttt{gengraph} routine uses to generate the amplitude. The
\texttt{FORM} model files are located in \texttt{models/form/} in the
user's working directory, and consist entirely of \texttt{FORM}
folds. None of these folds may be omitted from the file. We recommend
using an existing model file as a template for creating a new one.  An
empty template is provided in the \MaRTIn{} source directory at
{\texttt{user\_template/models/form/model\_MODEL}}. A full working
example, \texttt{model\_SM}, can be found in the same folder.

\subsubsection{Internal representation of propagators and vertices\label{sec:int:rep}}

The \texttt{gengraph} routine uses the raw \texttt{QGRAF} output to generate
a symbolic amplitude according to the Feynman rules implemented
in the \texttt{FORM} model file. The essential steps comprise the
generation of all group indices, insertion of the Feynman rules, and
ordering of the fermion lines. In this section, we briefly describe
how propagators and vertices are represented internally, and introduce
\MaRTIn{}'s fermionline objects.

All propagators and vertices are initially (i.e. before the
substitution of the Feynman rules) represented by commuting
functions. Their names are assigned by the \texttt{pfunct} and
\texttt{vfunct} statements in the \texttt{QGRAF} model files. These
commuting functions must be declared in the \foldface{DEF} fold of the
\texttt{FORM} model file (see Sec.~\ref{sec:other}). In addition,
commuting functions representing external leg factors are
automatically generated. Therefore, all field (and anti-field)
variables used in the \texttt{QGRAF} model files must also be declared as
symbols in the \foldface{DEF} fold.

Using \MaRTIn{}'s style file, \texttt{QGRAF} provides a unique
description of the topology of any given Feynman diagram by providing
either unique pairs of positive numbers, with one number being
associated with one end of a propagator and the same number being
associated with the corresponding (internal) leg of a vertex; or by
providing a unique pair of negative numbers, with one number being
associated with an external leg of a vertex, and the same number
labeling a corresponding ``polarization function''. The initial
symbolic representation of any propagator function \texttt{PROP} is,
thus, of the form
\begin{lstlisting}
    PROP(@$i_1$@,@$i_2$@,mom,@$p$@,mass,@$m$@,field,@$f$@)
\end{lstlisting}
Here, $i_1$ and $i_2$ are two unique positive integers provided by
\texttt{QGRAF}. They are followed by the momentum $p$ of the
propagator, the mass $m$ of the propagator, and a symbol $f$ denoting
the name of the field to which the propagator belongs. These variables
are indicated by the ``separator'' variables \texttt{mom},
\texttt{mass}, and \texttt{field}, respectively. These arguments help
the user identifying each propagator and inserting the correct Feynman
rule.

The situation is similar for all $N$-leg vertices. Here the order of
fields is the same as defined in the \texttt{QGRAF} model file. Any
vertex function \texttt{VERTEX} is of the form
\begin{lstlisting}
    VERTEX(@$i_1, \ldots, i_N$@,mom,@$p_1, \ldots, p_N$@)
\end{lstlisting}
where $i_1, \ldots, i_N$ are unique integers provided by
\texttt{QGRAF}, positive of negative depending on whether the
corresponding leg is internal or external. \texttt{mom} is a
separator, followed by the incoming momenta for each leg.

Finally, each external leg receives a polarization function
\texttt{POL} of the form
\begin{lstlisting}
    POL(@$f$@,@$i$@,@$p$@)
\end{lstlisting}
with $f$ again the symbol denoting the name of the corresponding
field, $i$ a unique negative integer provided by \texttt{QGRAF}, and
$p$ the momentum of the external line.  Whether the momentum is
incoming or outgoing depends on whether the corresponding leg/state
has been chosen as incoming or outgoing in the \foldface{QGRAF} fold.

These representations are the starting point of all further
manipulations, described in the following subsections.

\subsubsection{Generation of group indices}

While the gauge-group and global symmetry structure of the Feynman
rules is model-dependent, the Lorentz structure of the Feynman rules
is more generic. Thus, \MaRTIn{} will always generate Lorentz indices
automatically. Specifically, for each \texttt{pfunct} propagator
function \texttt{PROP}, \MaRTIn{} will generate a unique pair of
Lorentz indices as follows:
\begin{lstlisting}
    PROP(@$i_1$@,@$i_2$@,lorentz,@$\nu_{i_1}$@,@$\nu_{i_2}$@,mom,@$p$@,mass,@$m$@,field,@$f$@)
\end{lstlisting}
Similarly, for each \texttt{vfunct} vertex function \texttt{VERTEX}
corresponding to a vertex with $N$ legs, \MaRTIn{} will generate
Lorentz indices for each leg as follows:
\begin{lstlisting}
    VERTEX(@$i_1, \ldots, i_N$@,lorentz,@$\nu_{i_1}, \ldots, \nu_{i_N}$@,mom,@$p_1, \ldots, p_N$@)
\end{lstlisting}
The polarization functions are treated in the same way. In each case,
a new separator \texttt{lorentz} is generated, followed by $N$ unique
{\em internal} or {\em external} Lorentz indices of the form
\texttt{nu1} (internal) or \texttt{mu1} (external). The actual
integers appearing in these indices are in one-to-one correspondence
with the positive or negative integers generated by \texttt{QGRAF},
respectively. The Lorentz indices are used to generate the explicit
expression for the Feynman rules in the remainder of the code. (In
particular, they are just dropped for any scalar or fermionic
legs). {\em All other (gauge or global) indices need to be generated
  explicitly by the user in the model file.}

As an aid for the user, the module \texttt{docolour} is included in
the distribution of \MaRTIn{}; it may be used to handle fields
transforming under the fundamental or adjoint representations of an
$SU(N)$ algebra. (In particular, this comprises the case of QCD.) The
module handles both index generation and algebra, and provides the
following functions:\\[0.5ex]
\begin{tabular}{lll}
\texttt{Dcol(i,j)} & $\delta^{ij}$ & Kronecker delta for fundamental indices \\
\texttt{Dadcol(a,b)} & $\delta^{ab}$ & Kronecker delta for adjoint indices \\
\texttt{Tcol(i,j,a)} & $T_{ij}^{a}$ & generator of the fundamental representation \\
\texttt{Fcol(a,b,c)} & $f^{abc}$ & totally antisymmetric structure constants \\
\texttt{Cold(a,b,c)} & $d^{abc}$ & totally symmetric tensor
\end{tabular}\\[0.5ex]
These tensors are defined such that
$\text{Tr}\{T^a T^b\} = \tfrac{1}{2}\delta^{ab}$ and
$\text{Tr}\{T^a T^b T^c\} = \tfrac{1}{4}(d^{abc} + if^{abc})$.

These functions require external or internal leg numbers as arguments,
which will then be substituted in terms of unique indices by
\texttt{docolour}. For instance, group indices for a quark propagator
in the fundamental representation could be generated by the following
substitution:
\begin{lstlisting}
  id propagator(xx1?,xx2?,?a) = propagator(xx1,xx2,?a)*Dcol(xx1,xx2) ;
\end{lstlisting}
\texttt{docolour} then needs to be called to generate the appropriate
indices and subsequently resolve the $SU(N)$ algebra. Fundamental
indices are called \texttt{i1}, \texttt{i2}, etc. for internal legs
and \texttt{j1}, \texttt{j2}, etc. for external legs, and have
dimension \texttt{NcolF} ($=3$ for QCD), while adjoint indices are
called \texttt{a1}, \texttt{a2}, etc. for internal legs and
\texttt{b1}, \texttt{b2}, etc. for external legs, and have dimension
\texttt{NcolA} ($=8$ for QCD). It is recommended to include the
\texttt{\#include docolour} statement only after all group structures
are generated -- either directly at the end of the
\foldface{INSERTVERTICES} fold of the \texttt{FORM} model file (see
below), or within the \foldface{FOLD1} fold in the problem file.

Generation of all other indices is left to the user (for instance,
$SU(2)$ gauge indices for the SM in the unbroken phase). It is
recommended to use the leg numbers to generate all group indices. As
an example, consider the assignment of some flavour indices
\texttt{idx`i'} with dimension \texttt{Nf}. This can be done by
defining a \texttt{set}
\begin{lstlisting}
    Symbol Nf;
    Index idx1,...,idx99 = Nf;
    Set idcs: idx1,...,idx99;
\end{lstlisting}
which aids in matching indices to leg numbers later. For instance, a
propagator could be substituted by the following expression:
\begin{lstlisting}
    id propagator(xx1?pos_,xx2?pos_,?a) = d_(idcs[xx1],idcs[xx2])
                                          * propagator(xx1,xx2,?a);
\end{lstlisting}
External legs have negative indices and thus  need to be
handled separately.

The appropriate fold to generate group-theory structures is the
\foldface{GROUPTHEORY} fold in the \texttt{FORM} model file.
\begin{lstlisting}
  *--#[ GROUPTHEORY :
  @\textcolor{comment}{* ~ Possible place to factor out index contractions from propagators and vertices. }@

  @\textcolor{comment}{* ~ For instance, use the docolour module. }@
  @\textcolor{comment}{* ~ You only have to pass the leg numbers that QGRAF has assigned.  }@

  @\textcolor{comment}{* ~ Example: insert the SU(n) generator for the quark-gluon vertex.  }@
    id Vqqg(xx1?,xx2?,xx3?,?a) = Vqqg(xx1,xx2,xx3,?a)*Tcol(xx1,xx2,xx3);

  @\textcolor{comment}{* ~ Note that the index contractions may also be handled in later folds. }@
  *--#] GROUPTHEORY :
\end{lstlisting}

\subsubsection{Generation of fermion lines}

\MaRTIn{} does not use any explicit spinor indices for fermionic
variables, and instead uses its own as well as \texttt{FORM}'s
built-in expressions for Dirac matrices. This entails that all
fermionic variables (spinor functions and Dirac matrices) have to be
treated as non-commuting objects, and they have to be arranged in the
correct order. We will briefly describe the algorithm here, so the
user can implement their own fermionic vertices.

\texttt{QGRAF} requires that all fermionic vertices are bilinear in
the fermion fields. Therefore, fermionic vertices and propagators can
be used to construct \texttt{FL`i'} objects representing open and
closed fermion lines, enumerated by an integer \texttt{`i'}. These
objects represent products of Dirac matrices arising from both fermion
propagators and vertices in the correct order. After \texttt{gengraph}
has collected all fermion propagators and vertices into the
fermion-line objects, the corresponding propagator and vertex functions no
longer appear in the amplitude. Instead, each propagator is
represented as an argument inside \texttt{FL`i'}, of the form
\begin{lstlisting}
    FL1(?a,prop,mom,@$p$@,mass,@$m$@,?b)
\end{lstlisting}
Here, the separator \texttt{prop} indicates the presence of a
propagator factor; $p$ is any linear combination of loop and external
momenta flowing through the propagator, and $m$ is its
mass. Each vertex is identified by a symbol representing the vertex
type, followed by the arguments of the corresponding vertex
function. If \texttt{vertex} is the symbol associated with the vertex
function \texttt{VERTEX}, this vertex would be represented as follows:
\begin{lstlisting}
    FL1(?a,vertex,lorentz,@$\nu_{i_1}, \ldots, \nu_{i_N}$@,mom,@$p_1, \ldots, p_N$@,?b) * vertex
\end{lstlisting}
Note that the symbol \texttt{vertex} is used both as a label inside
the fermion line object (and will later be used by \texttt{FORM}'s
pattern matcher to substitute the vertex by the appropriate Dirac
structure), and as a symbol multiplying the term (this instance of the
symbol can later be replaced by some  value for the
corresponding Feynman rule, e.g, for the $\bar e\slashed{A} e$  vertex in QED
this could be $\texttt{vAee}\to i e Q_e$).

For the algorithm to work properly, all fermionic vertices must be
defined in the \texttt{QGRAF} model file with the antifermion field as
the first entry and the fermion field as the second entry. Moreover,
there must be a one-to-one correspondence between fermionic vertex
functions and the associated symbols. This is achieved by providing a
list of all fermionic vertices in the \foldface{FF} fold of the
\texttt{FORM} model file, and a list of all corresponding vertex
symbols in the \foldface{VERTICES} fold, {\em in exactly the same
  order}. This could look as follows:
\begin{lstlisting}
*--#[ FF :
@\textcolor{comment}{* ~ In order to generate spinor contractions, the vfunct of each vertex with }@
@\textcolor{comment}{* ~ fermions have to be listed here, comma separated, e.g.: }@
  Vqqg, Vqqs,
*--#] FF :

*--#[ VERTICES :
@\textcolor{comment}{* ~ For each vertex from the fold FF above, a symbol representing the coupling }@
@\textcolor{comment}{* ~ at this vertex has to be provided. The order has to match. }@
@\textcolor{comment}{* ~ MaRTIn will match these symbols to substitute the Feynman rules. }@
  vqqg, vqqs,
*--#] VERTICES :
\end{lstlisting}
where, in the example above, \texttt{Vqqg} and \texttt{Vqqs} are
declared as commuting functions and \texttt{vqqg} and \texttt{vqqs} as
symbols. Note that the trailing ``\texttt{,}'' is required. If any
vertex is not included in these lists, the code will exit with an
error message. If the ordering of functions and symbols is not in
exact correspondence, the Feynman rules will be inserted in a wrong
way. The actual substitution of the vertices by the Feynman rules is
discussed in Sec.~\ref{sec:fermion:vertices}.

\subsubsection{Substitution of propagators}

After the discussion of the fermion line objects, we next
describe the substitution of propagators. In general, the internal
function $\mathtt{Deno}(p,M) = 1/(p^2 - M^2)$ is employed to represent
the propagator denominators. The {\em numerators} $(\slashed{p}+m)$ of
all fermionic propagators are substituted automatically, in the
correct order, by \texttt{gengraph} using the fermion-line objects.
For this to work properly, a substitution of the following form is required in the
\foldface{INSERTPROPAGATORS} fold.
In the example below, \texttt{Q} is the quark-propagator function
  eventually corresponding to $i\delta^{ab}(\slashed{p}+m)/(p^2-m^2)$:
\begin{lstlisting}
  *--#[ INSERTPROPAGATORS :
  @\textcolor{comment}{* ~ Substitution of a quark propagator: }@
    id Q(xx1?,xx2?,lorentz,?x,mom,?p,mass,M?,field,fname?)
       = i_*F(xx1,xx2,mom,?p,mass,M,field,fname)*Deno(?p,M)*Dcol(xx1,xx2);

  @\textcolor{comment}{* ~ The internal function, \texttt{F}, must be used to properly account}@
  @\textcolor{comment}{* ~ for the $\gamma$-matrices and spinor structure of the numerator.}@
  @\textcolor{comment}{* ~ Note the use of the predefined commuting function Deno(p,M)~$ = 1/(p^2 - M^2)$. }@
  @\textcolor{comment}{* ~ Here, the colour factor has been supplied together with the denominator. }@
  *--#] INSERTPROPAGATORS :
\end{lstlisting}
The use of the internal \MaRTIn{} function, \texttt{F}, for the numerator part of
any fermionic propagator is essential for \texttt{gengraph} to work.

It remains to substitute all the bosonic propagators using the
\foldface{INSERTPROPAGATORS} fold:
\begin{lstlisting}
  *--#[ INSERTPROPAGATORS :
  @\textcolor{comment}{* ~ For instance, the substitution of a gluon propagator in $R_\xi$ gauge: }@
    id Dg(xx1?,xx2?,lorentz,nu1?,nu2?,mom,p?,mass,M?,field,g)
       = -i_*(d_(nu1,nu2) - p(nu1)*p(nu2)*Deno(p,M)*[1-xi])
          *Deno(p,M)*Dadcol(xx1,xx2);

  @\textcolor{comment}{* ~ [1-xi] is a symbol representing (one minus) the gauge-fixing parameter, i.e., $1-\xi$.}@
  *--#] INSERTPROPAGATORS :
\end{lstlisting}

\subsubsection{Substitution of fermionic vertices}\label{sec:fermion:vertices}

The substitution of fermionic vertices appearing in the \texttt{FL`i'}
objects can be done in two different ways. A large number of
frequently occurring vertices has been implemented in a generic way;
these vertices can be substituted by simply populating the
corresponding folds in the \texttt{FORM} model file with the vertex
symbols of the \foldface{VERTICES} fold. The following generic cases
are implemented:

{\bfseries Vector coupling:} For all vertex symbols \texttt{vv},
provided as a comma-separated list in the \foldface{VVCOUP} fold, the
Feynman rule $\gamma^\mu$ is substituted.
\begin{lstlisting}
  *--#[ VVCOUP :
  @\textcolor{comment}{* ~ For each vertex represented by the symbol listed here, }@
  @\textcolor{comment}{* ~ a vector-fermion-fermion will be susbtituted (e.g.: $\texttt{vv} \mapsto \texttt{vv} \,\gamma^\mu$.) }@
    vv, @\textit{...}@
  *--#] VVCOUP :
\end{lstlisting}
Note that after the substitution, the vertex is multiplied by the
corresponding vertex symbol. This symbol can then be used to insert
coupling constants or symmetry factors, see Sec.~\ref{sec:other}.

{\bfseries Chiral vector couplings:} For all vertex symbols
\texttt{va}, \texttt{vl}, \texttt{vr}, provided as three
comma-separated lists in the \foldface{VACOUPV}, \foldface{VACOUPL},
and \foldface{VACOUPR} folds, respectively, a Feynman rule
is substituted as follows:
\begin{lstlisting}
  @\textcolor{comment}{* ~ VACOUPV contains a list of vertex symbols, e.g. va, which must}@
  @\textcolor{comment}{* ~ also be contained in the \foldface{VERTICES} fold.}@
  @\textcolor{comment}{* ~ VACOUPL and VACOUPR contain corresponding lists of symbols, }@
  @\textcolor{comment}{* ~ vvl and vvr, for each VACOUPV symbol. }@
  @\textcolor{comment}{* ~ Thus, the number and order of elements must match in VACOUPV, VACOUPL, VACOUPR. }@
  @\textcolor{comment}{* ~ va gets replaced by vl, vr : $\texttt{va} \mapsto \gamma^\mu \,(\texttt{vl}\,P_L + \texttt{vr}\,P_R  )$. }@

  *--#[ VACOUPV :
    va,
  *--#] VACOUPV :
  *--#[ VACOUPL :
    vl,
  *--#] VACOUPL :
  *--#[ VACOUPR :
    vr,
  *--#] VACOUPR :

  @\textcolor{comment}{* ~ These three folds belong together and should have the same number of entries. }@
\end{lstlisting}
Here, $P_L = \tfrac{1}{2}(1-\gamma_5)$ and
$P_R = \tfrac{1}{2}(1+\gamma_5)$ denote the projectors onto the left-
and right-handed chiral states. Again, the symbols \texttt{vl} and
\texttt{vr} can be replaced later.

{\bfseries Scalar couplings:} For all vertex symbols \texttt{vs},
provided as a comma-separated list in the \foldface{SCOUPV} fold, a
unit matrix in spinor space $\mathbb{1}$ is substituted.
\begin{lstlisting}
  *--#[ SCOUPV :
  @\textcolor{comment}{* ~ Generic scalar-fermion-fermion operator, a unit matrix is substituted. }@
  @\textcolor{comment}{* ~ Same syntax as in the examples above: $\texttt{sv} \mapsto \texttt{sv}\,\mathbb{1}$ }@
  vs,
  *--#] SCOUPV :
\end{lstlisting}

{\bfseries Chiral scalar couplings:} For all vertex symbols
\texttt{sav}, \texttt{slv}, \texttt{srv}, provided as three
comma-separated lists in the \foldface{SACOUPV}, \foldface{SACOUPLV},
and \foldface{SACOUPRV} folds, respectively, a Feynman rule
is substituted as follows:\begin{lstlisting}
  @\textcolor{comment}{* ~ SACOUPV contains a list of vertex symbols; e.g. sav }@
  @\textcolor{comment}{* ~ SACOUPLV and SACOUPRV contain corresponding lists of symbols, }@
  @\textcolor{comment}{* ~ slv and srv, for each SACOUPV symbol. }@
  @\textcolor{comment}{* ~ Thus, the number and order of elements must match in SACOUPV, SACOUPLV, SACOUPRV. }@
  @\textcolor{comment}{* ~ sav gets replaced by slv, srv : $\texttt{sav} \mapsto \, \texttt{slv}\,P_L + \texttt{srv}\,P_R$. }@

  *--#[ SACOUPV :
    sav,
  *--#] SACOUPV :
  *--#[ SACOUPLV :
    slv,
  *--#] SACOUPLV :
  *--#[ SACOUPRV :
    srv,
  *--#] SACOUPRV :
\end{lstlisting}

This concludes the list of implemented fermion vertices. We would like
to remind the reader that, while folds may be empty, they should never
be removed from the model file.

Alternatively, fermion vertices can by substituted on a case-by-case
basis, using the fermion line objects. This has to be done in the
\foldface{INSERTFERMIONVERTICES} fold. All substitution rules provided
in this fold will by looped over the number of fermion line objects by
\texttt{gengraph}; the corresponding loop variable (fermion line
index) must be denoted as \texttt{`i'}. Care should be taken to
keep the correct ordering of the spinorial objects. For instance, the
vectorial quark-gluon interaction vertex could be implemented as
follows:
\begin{lstlisting}
  *--#[ INSERTFERMIONVERTICES :
  @\textcolor{comment}{* ~ Fermionic vertices, represented by fermion line (FL) objects}@
  @\textcolor{comment}{* ~ that are enumerated by `i', can be substituted here.  }@

  @\textcolor{comment}{* ~ Below is an example for the gluon - quark vertex, with vertex symbol vqqg. }@
    id FL`i'(?a,vqqg,lorentz,uu1?,mom,v1?,v2?,v3?)
       = FL`i'(?a) * g_(`i',uu1);
  *--#] INSERTFERMIONVERTICES :
\end{lstlisting}

An extended example showing how to implement an effective four-quark
vertex is included in App.~\ref{sec:four:quark}.

A few comments are in order. All Lorentz indices corresponding to the
fermion field indices are automatically deleted by \texttt{gengraph};
hence, the remaining Lorentz indices correspond to the fields
appearing after the two fermion fields in the definition of the vertex
in the \texttt{QGRAF} model file. In particular, the antifermion /
fermion fields must always appear at the first / second position in
the definition of the vertices in the \texttt{QGRAF} model files. The
preprocessor variable \texttt{`i'} enumerating the fermion line is
tracked also throughout the spinor matrices in the Feynman rule, which
can be \texttt{FORM}'s internal non-commuting functions for spinor
objects, such as the identity in spinor space (\texttt{gi\_}), the
$\gamma^\mu$ matrix (\texttt{g\_}), $\gamma_5$
(\texttt{g5\_}). Alternatively, \MaRTIn{}'s predefined objects
\texttt{PL`i'} and \texttt{PR`i'} for the chiral projectors and
\texttt{G5`i'} for the $\gamma_5$ matrix can be used.

If there are any remaining fermion line objects left after vertex
susbtitution, \MaRTIn{} will exit with an error message.

\subsubsection{Substitution of bosonic vertices}

As for fermionic vertices, there are several generic bosonic vertices
implemented that can be substituted in a simple manner via predefined
folds. They are listed in the following. For each of the implemented
cases, the first fold should contain the \texttt{vfunct} vertex
function corresponding to a given vertex, and the other folds should
contain corresponding user-defined symbols, as needed. These symbols
need to be declared in the \texttt{FORM} model file (see
Sec.~\ref{sec:other}). Again, we would like to remind the reader that,
while folds may be empty, they should never be removed from the model
file. Folds corresponding to a given vertex class need to be populated
with lists of functions or symbols of the same length, and in
one-to-one correspondence to each other.

{\bfseries Vector-vector interaction:} This vertex can be used, for
instance, to implement gauge-boson propagator counterterm
vertices. Here, $q$ is the incoming momentum of the first vector
field.
\begin{lstlisting}
  @\textcolor{comment}{* ~ Generic vector-vector vertex: $\texttt{Vv} \mapsto \texttt{vv1}\,q^2\,g^{\mu\nu} + \texttt{vv2}\,g^{\mu\nu} + \texttt{vv3}\,q^{\mu}q^{\nu}$. }@

  *--#[ VVgenCOUP :
    Vv,
  *--#] VVgenCOUP :
  *--#[ vVVgenCOUP1 :
    vv1,
  *--#] vVVgenCOUP1 :
  *--#[ vVVgenCOUP2 :
    vv2,
  *--#] vVVgenCOUP2 :
  *--#[ vVVgenCOUP3 :
    vv3,
  *--#] vVVgenCOUP3 :
\end{lstlisting}

{\bfseries Scalar-scalar interaction:} This vertex can be used, for
instance, to implement scalar propagator counterterm vertices. Here,
$q$ is the incoming momentum of the first scalar field.
\begin{lstlisting}
  @\textcolor{comment}{* ~ Generic scalar-scalar vertex: $\texttt{Ss} \mapsto \texttt{ss1}\,q^2 + \texttt{ss2}$. }@

  *--#[ SSgenCOUP :
    Ss
  *--#] SSgenCOUP :
  *--#[ vSSgenCOUP1 :
    ss1,
  *--#] vSSgenCOUP1 :
  *--#[ vSSgenCOUP2 :
    ss2,
  *--#] vSSgenCOUP2 :
\end{lstlisting}

{\bfseries Vector-scalar interaction:} This vertex can be used, for
instance, to implement vector-scalar mixing. Here, $q$ is the incoming
momentum of the vector field. Note that the vector field has to
precede the scalar field in the definition of the vertex in the
\texttt{QGRAF} model file.
\begin{lstlisting}
  @\textcolor{comment}{* ~ Generic vector-scalar (with momentum $q$) vertex: $\texttt{Vs} \mapsto \texttt{vs1}\,q^\mu$. }@

  *--#[ VSgenCOUP :
    Vs,
  *--#] VSgenCOUP :
  *--#[ vVSgenCOUP1 :
    vs1,
  *--#] vVSgenCOUP1 :
\end{lstlisting}

{\bfseries Triple vector interaction:} Here, $q_1$, $q_2$, $q_3$ are
the incoming momenta of the first, second, and third vector field,
respectively, while $\mu_1$, $\mu_2$, $\mu_3$ are their Lorentz
indices (ordering as in the \texttt{QGRAF} model file). Here and
below, $g^{\mu\nu}$ denotes the metric tensor of Minkowski space, and
$\epsilon_{ijk}$ is the totally antisymmetric Levi-Civita tensor in
three dimension with $\epsilon_{123}=1$.
\begin{lstlisting}
  @\textcolor{comment}{* ~ Generic triple vector vertex: $\texttt{Vvv} \mapsto \texttt{vvv} \,  \epsilon_{ijk}\, g^{\mu_i \mu_j} (q_j)^{\mu_k}$. }@

  *--#[ VVVgenCOUP :
    Vvv,
  *--#] VVVgenCOUP :

  *--#[ vVVVgenCOUP :
    vvv,
  *--#] vVVVgenCOUP :
\end{lstlisting}

{\bfseries Scalar-vector-vector interaction:} Here, $\mu$, $\nu$
denote the Lorentz indices of the two vector fields. The scalar field
must be at the first position in the \texttt{QGRAF} vertex definition.
\begin{lstlisting}
  @\textcolor{comment}{* ~ Generic scalar-vector-vector vertex: $\texttt{Svv} \mapsto \texttt{svv} \, g^{\mu \nu}$. }@

  *--#[ SVVgenCOUP :
    Svv,
  *--#] SVVgenCOUP :
  *--#[ vSVVgenCOUP :
    svv,
  *--#] vSVVgenCOUP :
\end{lstlisting}

{\bfseries Triple scalar interaction:}
\begin{lstlisting}
  @\textcolor{comment}{* ~ Generic triple scalar vertex: $\texttt{Sss} \mapsto \texttt{sss} $. }@

  *--#[ SSSgenCOUP :
    Sss,
  *--#] SSSgenCOUP :
  *--#[ vSSSgenCOUP :
    sss,
  *--#] vSSSgenCOUP :
\end{lstlisting}

{\bfseries Quartic scalar interaction:}
\begin{lstlisting}
  @\textcolor{comment}{* ~ Generic quartic scalar vertex: $\texttt{Ssss} \mapsto \texttt{ssss} $. }@

  *--#[ SSSSgenCOUP :
    Ssss,
  *--#] SSSSgenCOUP :
  *--#[ vSSSSgenCOUP :
    ssss,
  *--#] vSSSSgenCOUP :
\end{lstlisting}

{\bfseries Vector-vector-scalar-scalar interaction:} Here, $\mu$,
$\nu$ denote the Lorentz indices of the two vector fields. The two
vector fields must be at the first and second positions in the
\texttt{QGRAF} vertex definition.
\begin{lstlisting}
  @\textcolor{comment}{* ~ Generic vector-vector-scalar-scalar vertex: $\texttt{Vvss} \mapsto \texttt{vvss} \, g^{\mu \nu}$. }@

  *--#[ VVSSgenCOUP :
    Vvss,
  *--#] VVSSgenCOUP :
  *--#[ vVVSSgenCOUP :
    vvss,
  *--#] vVVSSgenCOUP :
\end{lstlisting}

{\bfseries Quartic vector interaction:} Here, $\mu_1,\ldots,\mu_4$
denote the Lorentz indices of the first, second, third, fourth vector
field as defined in the \texttt{QGRAF} model file.
\begin{lstlisting}
  @\textcolor{comment}{* ~ Generic quartic vector vertex: $\texttt{Vvvv} \mapsto \texttt{vvvv}  \, ( 2\,g^{\mu_1 \mu_2}g^{\mu_3 \mu_4} - g^{\mu_1 \mu_3}g^{\mu_2 \mu_4} - g^{\mu_1 \mu_4}g^{\mu_2 \mu_3})$. }@

  *--#[ VVVVgenCOUP :
    Vvvv,
  *--#] VVVVgenCOUP :
  *--#[ vVVVVgenCOUP :
    vvvv,
  *--#] vVVVVgenCOUP :
\end{lstlisting}

{\bfseries Vector-scalar-scalar interaction:} Here, $\mu_1$ is the
Lorentz index of the vector field (first field in the \texttt{QGRAF}
model file), while $q_2$, $q_3$ are the incoming momenta of the first
and second scalar field (second and third fields in the \texttt{QGRAF}
model file).
\begin{lstlisting}
  @\textcolor{comment}{* ~ Generic vector-scalar-scalar vertex: $\texttt{Vss} \mapsto \texttt{vss}  \, (q_2 - q_3)^{\mu_1}$. }@

  *--#[ VSSgenCOUP :
    Vss,
  *--#] VSSgenCOUP :
  *--#[ vVSSgenCOUP :
    vss,
  *--#] vVSSgenCOUP :
\end{lstlisting}

{\bfseries Vector-ghost-ghost interaction:} Here, $\mu_1$ is the
Lorentz index of the vector field (first field in the \texttt{QGRAF}
model file), while $q_2$ is the incoming momentum of the first ghost
field (second field in the \texttt{QGRAF} model file).
\begin{lstlisting}
  @\textcolor{comment}{* ~ Generic vector-ghost-ghost vertex: $\texttt{Vhh} \mapsto \texttt{vhh}  \, (q_2 )^{\mu_1}$. }@

  *--#[ VGGgenCOUP :
    Vhh,
  *--#] VGGgenCOUP :
  *--#[ vVGGgenCOUP :
    vhh,
  *--#] vVGGgenCOUP :
\end{lstlisting}

Alternatively, any bosonic vertices may be substituted on a
case-by-case basis in the \foldface{INSERTVERTICES} fold.
This may be required if other Lorentz structures appear in the Feynman rules, for
instance, when implementing Feynman rules in a background field gauge,
or in an effective field theory. The corresponding substitutions can
be constructed from the arguments of the vertex functions, discussed
in Sec.~\ref{sec:int:rep}. An example is given in the following.
\begin{lstlisting}
  *--#[ INSERTVERTICES :
  @\textcolor{comment}{* ~ Use arguments of vertex functions to substitute Feynman rules. }@
  @\textcolor{comment}{* ~ The order of arguments correspond to the order }@
  @\textcolor{comment}{* ~ of how fields were defined in QGRAF. }@
  @\textcolor{comment}{* ~ For convenience, the arguments are separated by keywords lorentz and mom.}@

  @\textcolor{comment}{* ~ Consider for instance the three-gluon-vertex: }@
    id Vggg(?a,lorentz,nu1?,nu2?,nu3?,mom,v1?,v2?,v3?) =
      v3g * (
        +d_(nu1,nu3)*(v3(nu2)-v1(nu2))
        +d_(nu2,nu3)*(v2(nu1)-v3(nu1))
        +d_(nu1,nu2)*(v1(nu3)-v2(nu3))
      );
  @\textcolor{comment}{* ~ It is assumed that the colour structure was generated before.}@
  @\textcolor{comment}{* ~ v3g is a symbol representing the coupling constant. }@
  *--#] INSERTVERTICES :
\end{lstlisting}

\subsubsection{Substitution of external leg factors}

Finally, external leg factors are substituted in the
\texttt{POLARIZATION} fold. In addition to the polarization function
discussed in Sec.~\ref{sec:int:rep}, \MaRTIn{} generates a
non-commuting
\begin{lstlisting}
  dummyspinor(fermionlineindex, fieldindex, dualfieldindex)
\end{lstlisting}
function for each external fermion. An example of how these objects
can be translated into meaningful external leg factors, i.e., the
external spinor functions $u(p)$ and $v(p)$,
is found in \texttt{model\_SM}. A simple
substitution rule for external photons could look like
\begin{lstlisting}
  *--#[ POLARIZATION :
  @\textcolor{comment}{* ~ Handle substitutions of polarization / dummyspinor functions here. }@
  @\textcolor{comment}{* ~ Below is example for a scalar field. }@
  @\textcolor{comment}{* ~ First we mark external leg numbers with nneg(): }@

    id pol(xx1, xx2?neg_, q1?) = pol(xx1, nneg(-xx2), q1);

  @\textcolor{comment}{* ~ which can be used to build polarisation vectors, e.g. for photons (field symbol ``a'')}@

    id pol(a, nneg(xx2?odd_), q1?) = ee(a, mom, q1, lorentz, mmuu[xx2]);
    id pol(a, nneg(xx2?even_), q1?) = eestar(a, mom, q1, lorentz, mmuu[xx2]);

  @\textcolor{comment}{* ~ distinguishing incoming and outgoing photon legs. }@
  @\textcolor{comment}{* ~ Here, ``mmuu'' is the built-in set of external Lorentz indices. }@
  *--#] POLARIZATION :
\end{lstlisting}
External fermions, as well as particles with additional quantum
numbers, can be handled in the same way; see the supplied
\texttt{model\_SM} for examples. It is highly recommended to generate
any spinor functions only in the \foldface{POLARIZATION} fold, to
avoid interference with the internal routines that handle the Dirac
algebra.

\subsubsection{Other folds\label{sec:other}}

Each \texttt{FORM} model file contains three \texttt{FORM} folds in addition to
those discussed above.

The \foldface{DEF} fold contains all \texttt{FORM} definitions
necessary for the implementation of the model. This includes all
fields (as symbols), propagator and vertex functions (commuting
functions) referenced in the \texttt{QGRAF} model file, but also
anything else the user desires. Note that many variables are already
defined in \texttt{prc/maindeclare}.
\begin{lstlisting}
  *--#[ DEF :
  @\textcolor{comment}{* ~ This fold contains all necessary declarations for the model file. }@

  @\textcolor{comment}{* ~ Here is a good place to define the integer spacetime dimension.}@
  @\textcolor{comment}{* ~ \MaRTIn{} has been most tested for the case of ``4'' and less so for ``3''.}@
    #define DIMENSION "4"

  @\textcolor{comment}{* ~ Declare all your symbols, functions, indices and sets here. }@
  @\textcolor{comment}{* ~ Several variables are already defined in prc/maindeclare.  }@
  @\textcolor{comment}{* ~ Fields and masses from the QGRAF model should be declared as symbols. }@
  @\textcolor{comment}{* ~ All pfunct and vfunct should be declared as commuting functions. }@

  *--#] DEF :
\end{lstlisting}

The \foldface{AMASSES} fold defines a \texttt{FORM} set that fixes the
hierarchy of masses in a given model, as well as a dollar variable set
to the total number of masses. This is used to reduce the number of
terms produced in some two-loop integrals when there are more than
three masses present.
\begin{lstlisting}
  *--#[ AMASSES :
  @\textcolor{comment}{* ~ In this fold we need to define the hierarchy of masses. }@
  @\textcolor{comment}{* ~ This information is required by some two-loop master integrals. }@

  @\textcolor{comment}{* ~ Fill this set with ordered masses. For M1 < M2 < M3 < M4: }@
    set massorder:  M1, M2, M3, M4;

  @\textcolor{comment}{* ~ The length of massorder needs to be stored in a dollar variable. }@
    #@$\texttt{\$}$@massnum = 4;

  *--#] AMASSES :
\end{lstlisting}

Finally, vertex symbols may be replaced with actual couplings in the
fold \foldface{INSERTCOUPLINGS}.
\begin{lstlisting}
  *--#[ INSERTCOUPLINGS :
  @\textcolor{comment}{* ~ At the very end of the calculation, insert the actual couplings. }@
  @\textcolor{comment}{* ~ Renormalisation may be implemented here. }@
    id vgqq = g + dg1/ep;

  *--#] INSERTCOUPLINGS :
\end{lstlisting}

\subsubsection{Summary}

Most of the information contained in the \texttt{FORM} model file
feeds into the internal module \texttt{gengraph}, which handles the
insertion of Feynman rules, see Sec.~\ref{sec:algo}. We close this
section by giving a graphical overview of \texttt{gengraph}'s
workflow, as the order of inclusion for each fold may be relevant to
the user.

\begin{tikzpicture}[node distance=1.5cm and 1.5cm]
  \node (gengraph) [rect, align=center] {\texttt{gengraph}:\\ set $\mathtt{q0} = \mathtt{M0} = 0$ \\ generate Lorentz indices, momenta};
  \node (expr) [input, right=1cm of gengraph, align=center] { diagram  expressions  \\from \texttt{QGRAF}};
  \node (grouptheory) [rect, align=center, below=0.5cm of gengraph] {splice out algebra \\ (optional)};
  \node (fold_grouptheory) [input, right=2.6cm of grouptheory, align=center] {\foldface{GROUPTHEORY} \\ fold in \texttt{model*.frm}};
  \node (propagators) [rect, align=center, below=0.5cm of grouptheory] {insert propagators \\ without spinor structure };
  \node (fold_propagators) [input, right=2.0cm of propagators, align=center] {\foldface{INSERTPROPAGATORS} \\ fold in \texttt{model*.frm}};
  \node (fli) [rect, align=center, below=0.5cm of propagators] {generate fermion \\ line objects };
  \node (cll) [input, right=2.6cm of fli, align=center] {\foldface{FF}, \foldface{VERTICES} folds in \texttt{model*.frm}\\ \foldface{CLLABEL} option in \texttt{loop.*.dat}};
  \node (finsert) [rect, align=center, below=0.7cm of fli] {insert manual spinor structures,\\  insert generic fermion vertices, \\ insert fermion propagator spinors};
  \node (ffolds) [input, right=1.1cm of finsert, align=center] {\foldface{INSERTFERMIONVERTICES}, \\ \foldface{VVCOUP}, \foldface{VACOUPV},\\ \foldface{VACOUPL}, \foldface{VACOUPR}, \\ \foldface{SACOUPV},  \foldface{SACOUPLV}, \foldface{SACOUPRV}\\ folds in \texttt{model*.frm}};
  \node (vinsert) [rect, align=center, below=4.2cm of finsert] {insert user-defined bosonic vertices, \\
  insert generic bosonic vertices
  };
  \node (vfolds) [input, align=center, right=0.9cm of vinsert] {
    \foldface{INSERTVERTICES}, \\
  \foldface{VVgenCOUP}, \foldface{vVVgenCOUP1}, \\  \foldface{vVVgenCOUP2}, \foldface{vVVgenCOUP3}, \\
  \foldface{SSgenCOUP}, \foldface{vSSgenCOUP1}, \\ \foldface{vSSgenCOUP2},
  \foldface{VSgenCOUP}, \\ \foldface{vVSgenCOUP1}, \foldface{VVVgenCOUP},  \\ \foldface{vVVVgenCOUP},
  \foldface{SVVgenCOUP},  \\ \foldface{vSVVgenCOUP},  \foldface{SSSgenCOUP}, \\ \foldface{vSSSgenCOUP},
  \foldface{SSSSgenCOUP}, \\ \foldface{vSSSSgenCOUP}, \foldface{VVSSgenCOUP},  \\ \foldface{vVVSSgenCOUP},
  \foldface{VVVVgenCOUP}, \\ \foldface{vVVVVgenCOUP}, \foldface{VSSgenCOUP}, \\ \foldface{vVSSgenCOUP},
  \foldface{VGGgenCOUP}, \\ \foldface{vVGGgenCOUP} \\
  folds in \texttt{model*.frm}};

  \draw [arrow] (expr) -- (gengraph);
  \draw [arrow] (fold_grouptheory) -- (grouptheory);
  \draw [arrow] (gengraph) -- (grouptheory);
  \draw [arrow] (fold_propagators) -- (propagators);
  \draw [arrow] (grouptheory) -- (propagators);
  \draw [arrow] (propagators) -- (fli);
  \draw [arrow] (cll) -- (fli);
  \draw [arrow] (fli) -- (finsert);
  \draw [arrow] (ffolds) -- (finsert);
  \draw [arrow] (finsert) -- (vinsert);
  \draw [arrow] (vfolds) -- (vinsert);

\end{tikzpicture}

\subsection{Add-ons for \MaRTIn{}}
Additional information on the models might be required for add-ons to \MaRTIn{}.

\begin{figure}[h!]
  \centering
  \includegraphics[scale=.9]{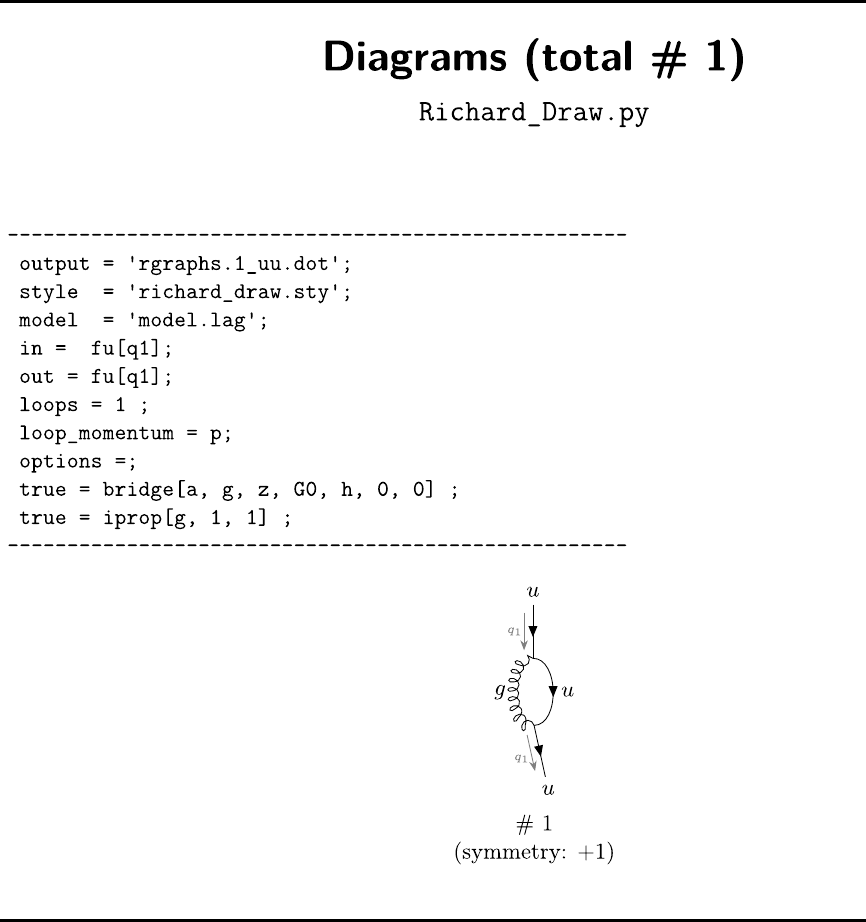}
  \caption{Example of \texttt{richard\_draw} output for the problem \texttt{SM/loop.1\_uu.dat}. }
  \label{fig:rdraw_example}
\end{figure}

Currently, this is only the case for \texttt{richard\_draw}.
This piece of software requires a model file \texttt{models/rdraw/\textit{qmodel}.json} of the working directory, where \texttt{\textit{qmodel}} has to match the name of the \texttt{QGRAF} files.
This JSON formatted file contains the following style information:
\begin{lstlisting}
  {
    "fields": {
      "@\textit{QGRAF name}@": ["@\textit{TEX name}@", "@\textit{linetype}@"],
      @\textit{...}@
    }
  }
\end{lstlisting}
where \texttt{"\textit{QGRAF name}"} has to match the field identifier in the \texttt{QGRAF}, \texttt{"\textit{TEX name}"} contains valid \LaTeX{} code, and \texttt{"\textit{linetype}"} controls the appearance of the propagator lines. The default options for the latter include \texttt{"fermion"}, \texttt{"anti fermion"}, \texttt{"scalar"}, \texttt{"charged scalar"}, \texttt{"anti charged scalar"}, \texttt{"boson"}, \texttt{"gluon"} and \texttt{"ghost"}.
An example file is provided by \texttt{user\_template/models/rdraw/standardmodel.json} in the source directory.
\MaRTIn{} may execute \texttt{richard\_draw} via the \texttt{rpdf} target, e.g. the command
\begin{lstlisting}
    martin problems/SM/loop.1_uu.dat rpdf
\end{lstlisting}
will generate the file displayed in Fig.~\ref{fig:rdraw_example}.

\section{Example\label{sec:Examples}}

In this section we briefly discuss the calculation of the divergent
part of the QCD quark self energy, in order to illustrate the usage of
\MaRTIn{}. This example uses the following files that are distributed
together with the code: The Standard Model \texttt{FORM}-model file
\texttt{model\_SM}, located in \texttt{user\_template/models/form}, the
Standard Model \texttt{qgraf}-model files
\texttt{standardmodel.prop.lag} and \texttt{standardmodel.vrtx.lag}, located in
\texttt{user\_template/models/qgraf}, and the problem file
\texttt{loop.1\_uu.dat} located in
\texttt{user\_template/problems/SM}. The latter is calling the
\texttt{dolour} procedure from the base code to simplify the colour
structure. This example should give a good illustration of how to read
the output of \MaRTIn{}.

The corresponding call of \MaRTIn{}, e.g., from the
\texttt{user\_template} directory, would be:
\begin{lstlisting}
martin problems/SM/loop.1_uu.dat
\end{lstlisting}
which would generate and compute all diagrams. In this example there is
a single diagram, see Fig.~\ref{fig:rdraw_example}.

This will result in a screen printout, which includes the following lines:
\begin{lstlisting}

-----------------------------------------------------------------------
Massive Recursive Tensor Integration

            M a R T I n

Copyright (C) 2009-2023 J. Brod, L. Huedepohl, E. Stamou, T. Steudtner
MaRTIn is licensed under the GNU General Public License Version 3.
-----------------------------------------------------------------------

Extracted default configuration from config file: "~/.martin.conf"

Computing using the input of the loop.dat:
   xxx/user_template/problems/SM/loop.1_uu.dat

The problem suffix of the loop.dat is:
   suffix = ".1_uu"

The problem directory is:
   xxx/user_template/problems/SM

The results directory is:
   [inferred from loop.dat] : xxx/user_template/results/SM

The user_prc directories are (in order of importance):
   [inferred from loop.dat] : xxx/user_template/prc

The FORM models are:
   [inferred from loop.dat] : xxx/user_template/models/form/model_SM

The QGRAF models are:
   [inferred from loop.dat] : xxx/user_template/models/qgraf/sm.prop.lag
   [inferred from loop.dat] : xxx/user_template/models/qgraf/sm.vrtx.lag

The JSON files for richard_draw.py are:
   [inferred from loop.dat] : xxx/user_template/models/rdraw/standardmodel.json

Computing based on GIT commit: "<hash>" (Modulo local/uncommited modifications)
\end{lstlisting}
These lines just explicitly list the files used by \MaRTIn{} in that
particular run. (In practice, \texttt{xxx} will correspond to the base
directory that is actually used.) The git commit version is also
indicated. This is followed by information on the generation of the
diagrams:
\begin{lstlisting}
Running "make"...

Generating xxx/user_template/results/SM/qgraf.1_uu.dat ...
Generating xxx/user_template/results/SM/qlist.1_uu.dat ...
Running QGRAF ...

  -------------------------------------------------------
                        qgraf-3.1.4
  -------------------------------------------------------

 output = 'qlist.1_uu.dat';
 style  = 'main.sty';
 model  = 'model.lag';
 in =  fu[q1];
 out = fu[q1];
 loops = 1 ;
 loop_momentum = p;
 options =;
 true = bridge[a, g, z, G0, h, 0, 0] ;
 true = iprop[g, 1, 1] ;

  -------------------------------------------------------

   24P  ---  7+  17-  ---  5N+  2C+  17C-

   163V  ---  3^136  4^27

  -------------------------------------------------------

    -   4^1  ---  0 diagrams
   3^2   -   ---  1 diagram

   total =  1 diagram

Generating xxx/user_template/results/SM/make.1_uu.info ...
rm xxx/user_template/results/SM/qlist.1_uu.dat
Generating xxx/user_template/results/SM/qgraf.1_uu.dat ...
Generating xxx/user_template/results/SM/form.1_uu.dat ...
\end{lstlisting}
Finally, \texttt{FORM} is invoked to perform the actual calculation:
\begin{lstlisting}
Computing xxx/user_template/results/SM/form.1_uu/dia1.sav ...
FORM 4.2.1 (Feb  6 2019, v4.2.1-3-g558b01f) 64-bits  Run: <date and time>
    #-

   dia1 =

       + ep^-1 * (
          - 3/4*UbarSp(fu,su3col,j1,mom,q1)*DIRAC(1,one)*USp(fu,su3col,j1,mom,
         q1)*i_*pi^-1*alphas*Mup*CF
          - 1/4*UbarSp(fu,su3col,j1,mom,q1)*DIRAC(1,one)*USp(fu,su3col,j1,mom,
         q1)*i_*pi^-1*xiqg*alphas*Mup*CF
          + 1/4*UbarSp(fu,su3col,j1,mom,q1)*DIRAC(1,q1)*USp(fu,su3col,j1,mom,
         q1)*i_*pi^-1*xiqg*alphas*CF
          );

  0.10 sec out of 0.10 sec
Done computing xxx/user_template/results/SM/form.1_uu/dia1.sav.

MaRTIn finished.
\end{lstlisting}
The following code in the \texttt{FORM} folds in the problem file has
been used to simplify the output
\begin{lstlisting}
*--#[ USERDEF :
s CF,alphas;
*--#] USERDEF :

*--#[ FOLD1 :
#include docolour
*--#] FOLD1 :

*--#[ FOLD5 :
id eM(mIRA) = 1;
id nc = 2*CF + 1/nc;
id gs^2 = 4*pi*alphas;
*--#] FOLD5 :
\end{lstlisting}
This is used to replace \texttt{eM(mIRA)} by $1$ and \texttt{gs} by
$\sqrt{4\pi\alpha_s}$, as well as powers of $N_c$ in terms of
$C_F \equiv (N_c^2-1)/2N_c$. Note that the spinorial objects
\texttt{UbarSp}, \texttt{DIRAC}, and \texttt{USp} are
non-commuting. The final output then corresponds to the familiar
expression
\begin{equation}
  \texttt{dia1} =
  - \frac{i\alpha_s}{4\pi} C_F \frac{1}{\varepsilon}
\bar u(\pmb{q}_1,j_1) \Big[ \xi_G \slashed{q}_1 +  m_u (3+\xi_G) \Big]u  (\pmb{q}_1,j_1) \,.
\end{equation}
The result is saved as \texttt{dia1.sav} in the directory
\texttt{user\_template/results/SM/form.1\_uu} and can be loaded using
\texttt{FORM}'s \texttt{load} command. A \texttt{Mathematica}-readable
(text-)file \texttt{dia1.m} is saved in the directory
\texttt{user\_template/results/SM/math.1\_uu}. Its contents are
\begin{lstlisting}
(* dia1 *)      { - 3/4*MathDirac[UbarSp[fu,su3col,j1,mom,q1],USp[fu,su3col,
      j1,mom,q1]]*I*Pi^(-1)*ep^(-1)*alphas*Mup*CF - 1/4*MathDirac[
      UbarSp[fu,su3col,j1,mom,q1],USp[fu,su3col,j1,mom,q1]]*I*Pi^(-1)*
      ep^(-1)*xiqg*alphas*Mup*CF + 1/4*MathDirac[UbarSp[fu,su3col,j1,
      mom,q1],q1,USp[fu,su3col,j1,mom,q1]]*I*Pi^(-1)*ep^(-1)*xiqg*
      alphas*CF}
\end{lstlisting}

\subsection*{Acknowledgements}

We would like to thank Martin Gorbahn for many cross checks of our
code during various fruitful collaborations. J.B. acknowledges support
in part by DoE grant DE-SC1019775.

\appendix

\section{Implementation of four-fermion vertices\label{sec:four:quark}}

\texttt{QGRAF} does not allow for the generation of local four-fermion
vertices. They can, however, be implemented in a straightforward way
using auxiliary propagators. The following example shows how to
implement the four-quark operator
$Q = (\bar s_L \gamma^\mu c_L)(\bar u_L \gamma_\mu d_L)$. Here,
$q_L \equiv \tfrac{1}{2}(1-\gamma_5)q$ denote the left-handed quark
fields ($q=u,d,s,c$).

The \texttt{QGRAF} model file should include
\begin{lstlisting}
  @\textcolor{comment}{* ~ Artificial propagator: }@
  [sp,sm,+;pfunct='Dmu',m='M0']

  @\textcolor{comment}{* ~ Vertices for the two quark bilinears: }@
  [fS,fc,sp;vfunct='FSc']
  [fU,fd,sm;vfunct='FUd']
\end{lstlisting}
We assume that all functions and symbols have been properly declared,
and color indices have been generated. The vertex can then be
substituted in two steps. If vertex symbols \texttt{vFSc} and
\texttt{vFUd} (together with the corresponding vertex functions
\texttt{FSc} and \texttt{FUd}) have been defined and included in the
\foldface{FF} and \foldface{VERTICES} folds, we include the following
substitution rule in the \foldface{INSERTFERMIONVERTICES} fold:
\begin{lstlisting}
*--#[ INSERTFERMIONVERTICES :
*

id FL`i'(?a,vFUd,lorentz,mu1?,mom,q1?,q2?,q3?)
   = FL`i'(?a)*g_(`i',mu1)*PL`i';

id FL`i'(?a,vFSc,lorentz,mu1?,mom,q1?,q2?,q3?)
   = FL`i'(?a)*g_(`i',mu1)*PL`i';

*
*--#] INSERTFERMIONVERTICES :
\end{lstlisting}
Here, we made use of the unique Lorentz indices that have been
generated automatically for the artificial ``propagator field''. (This
trick does not work for operators with longer strings of Dirac
matrices, in which case the Lorentz indices have to be inserted/generated ``by
hand''.)  Finally, we use the substitution of the artificial
propagator to contract the correct pairs of Lorentz indices, by simply
replacing it by a metric tensor:
\begin{lstlisting}
*--#[ INSERTPROPAGATORS :
*

id Dmu(xx1?,xx2?,lorentz,mu1?,mu2?,mom,?p,mass,M?,field,fname?)
   = d_(mu1,mu2) ;

*
*--#] INSERTPROPAGATORS :
\end{lstlisting}
Note that \texttt{gengraph} will automatically supply a factor
\texttt{vFSc*vFUd}, which can then be substituted by the Wilson
coefficient of the operator. For instance, if the Lagrangian in the
current example is ${\mathcal L} = \sqrt{2} G_F C Q$, one would
include the following substitution rule in the
\foldface{INSERTCOUPLINGS} fold:
\begin{lstlisting}
  *--#[ INSERTCOUPLINGS :

    id vFSc*vFUd = i_*sqrt_(2)*GF*C ;

  *--#] INSERTCOUPLINGS :
\end{lstlisting}

\cleardoublepage
\phantomsection
\addcontentsline{toc}{section}{References}
\begin{bibliography}{manual}
  \bibliographystyle{./bibstyles/manual}
\end{bibliography}

\end{document}